\documentclass[aps,prl,showpacs,11pt,onecolumn,notitlepage]{revtex4-1}
\usepackage{caption,bm,epsfig,graphics,amssymb,amsmath,subeqnarray,setspace,graphicx,amsthm,epstopdf,subfigure,color}

\def\b{\mathbf}\def\e{\epsilon}

\begin{document}
\title{Comparative hydrodynamics of bacterial polymorphism}
\author{Saverio E. Spagnolie}
\email{sespagnolie@ucsd.edu}
\author{Eric Lauga}
\email{elauga@ucsd.edu}
\affiliation{Department of Mechanical and Aerospace Engineering, University of California San Diego, 9500 Gilman Drive, La Jolla CA 92093-0411.}


\date{\today}

\begin{abstract}
Most bacteria swim through fluids by rotating helical flagella which can take one of twelve distinct polymorphic shapes. The most common helical waveform is the ``normal'' form, used during forward swimming runs. To shed light on the prevalence of the normal form in locomotion, we gather all available experimental measurements of the various polymorphic forms and compute their intrinsic hydrodynamic efficiencies. The normal helical form is found to be the most hydrodynamically efficient of the twelve polymorphic forms by a significant margin -- a conclusion valid for  both the peritrichous and polar flagellar families, and robust to a change in the effective flagellum diameter or length. The hydrodynamic optimality of the normal polymorph suggests that, although energetic costs of locomotion are small for bacteria,  fluid mechanical forces may have played a significant role in the evolution of the flagellum.
\end{abstract}

\maketitle

The shapes and sizes of life in all its diversity are ever changing as form meets function, intimately tuned to nature's diverse environments. Bacteria evolved to swim through fluids by rotating a single helical flagellum (``monotrichous'', or polar, bacteria), or in the case of such organisms as {\it Salmonella} and {\it Escherichia coli}, several rotating helical flagella emanating from their cell membranes (``peritrichous'' bacteria). Each flagellum is assembled through the polymerization of a flagellin protein, and has been met with great interest both in and outside the scientific community due to its astoundingly complex construction \cite{nv97}. Due to the various possible arrangements of polymerized flagellin, it has been postulated that the flagellar filaments can take only twelve distinct polymorphic forms \cite{Asakura70,Calladine78,hyn98}, of which nine have been characterized experimentally \cite{kay80} (Fig.~\ref{Figure1}a).

The most common helical waveform is the left-handed ``normal'' form, used during forward swimming ``runs.'' Upon counterclockwise (CCW [when viewed from the flagellum's distal end]) co-rotation of the flagella by rotary motors, a flagellar bundle forms behind peritrichous bacteria, driving fluid backward and propelling the cell forward. To change their swimming directions, these bacteria undergo ``tumbling'' events. As shown in Fig.~\ref{Figure1}b-e, a quick direction reversal to clockwise (CW) motor rotation produces a twisting torque which temporarily transforms the associated individual flagellum from a left-handed normal form to a right-handed ``semi-coiled'' form, leading to an unwinding of the bundle and a change in cell orientation, followed by a transition to a right-handed ``curly'' form which persists until the next reversal in motor direction \cite{Berg2003,db07,dtrb07}. The other forms are not generally used for locomotion.

\begin{figure}[b]
\begin{center}
\includegraphics[width=0.5\textwidth]{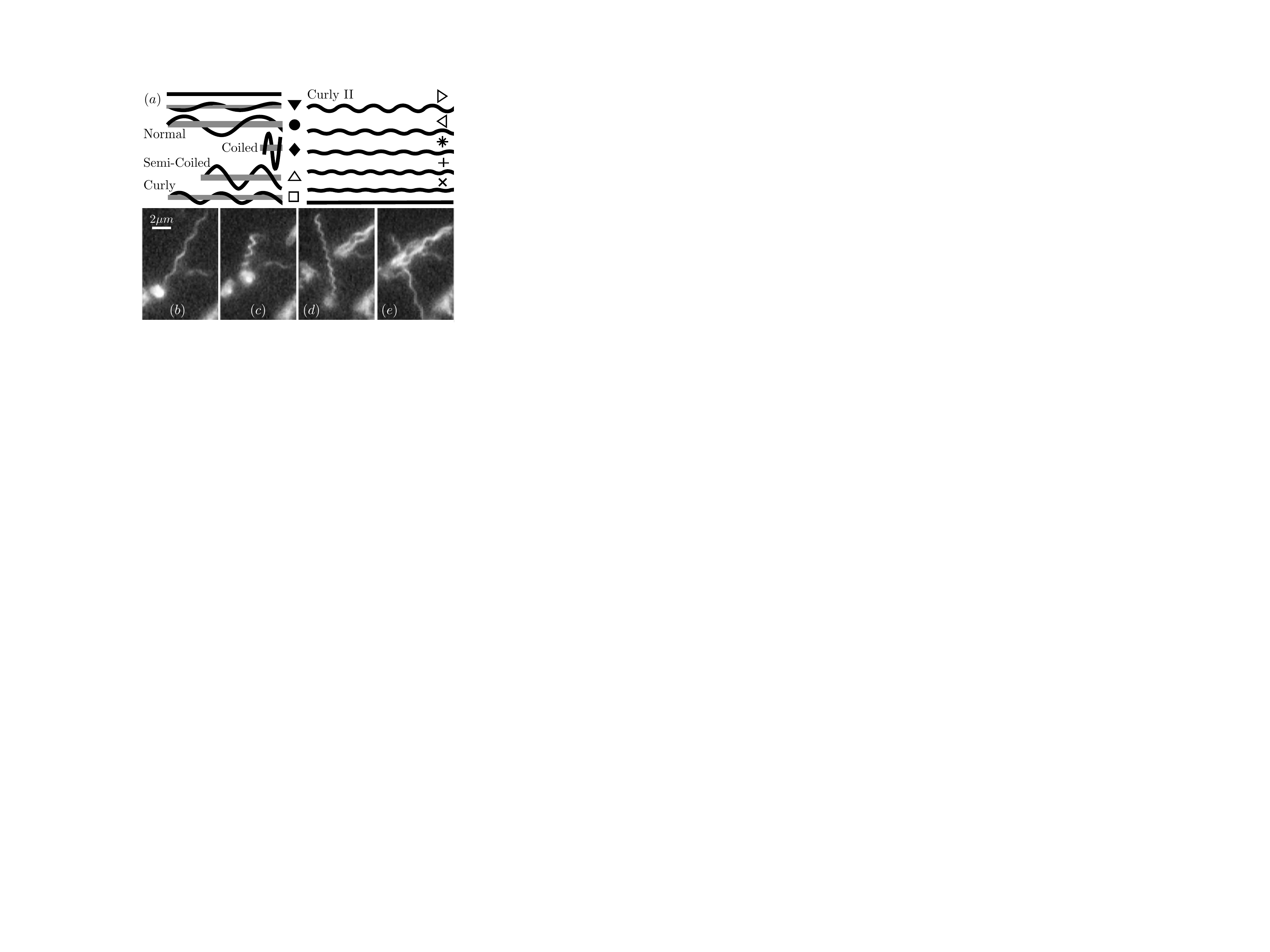}
\caption{FIG. 1 (a) All twelve theoretical peritrichous polymorphic waveforms, including two straight forms \cite{Calladine78}; left-handed (resp.~right-handed) helices are denoted by filled (resp.~empty) symbols. (b) One flagellum of an {\it E. coli} cell displays a normal waveform; (c) semi-coiled; (d) curly; (e) normal again. Adapted with permission from Turner, L., Ryu, W.S., and Berg, H.C., {\it J. Bacteriol.}, \textbf{182} 2793 (2000). Copyright  \copyright\,\,(2000), American Society for Microbiology \cite{twb00}.}
\label{Figure1}
\end{center}
\vspace{-.25in}
\end{figure}

Mechanical stresses, such as the twisting and viscous torques present during swimming, are not the only means by which the flagellar shape might shift from one waveform to another. Filaments can also transform reversibly in response to amino acid replacements, chemical or temperature changes, or the addition of alcohols or sugars \cite{Leifson60,ka76,mo77,Hotani80,hka82,Hotani82,ht91,sih93}. Other authors have considered the elastic rigidity of different polymorphs and its relationship to shape selection \cite{th92,gav06,sp06,db07}. The motion of a helical body through a viscous fluid has seen extensive theoretical treatment \cite{Taylor52,cw71,Lighthill75,Higdon79,Lighthill96}.

In this letter we present a physical rationalization of the  prevalence of the normal polymorphic form in bacterial swimming. We gather all available experimental measurements of the various polymorphic flagellar waveforms \cite{SuppMat} along with the twelve theoretical forms \cite{Calladine78}, and compute the intrinsic hydrodynamic efficiency of each geometry. We show that  the normal form is the most hydrodynamically efficient of the twelve polymorphic forms by a significant margin, a result true for both peritrichous and monotrichous (polar) flagellar families. This conclusion is robust as the flagellum   length is varied, or its effective diameter is increased to represent a bundle of flagella. The hydrodynamic optimality of the normal helical form therefore suggests a role for fluid mechanical forces in the evolution of the flagellum.

We begin with a short description of the hydrodynamics of swimming bacteria. At the exceedingly small length and velocity scales on which bacteria swim, viscous dissipation overwhelms any inertial effects, and the fluid motion is accurately described by the Stokes equations \cite{hb65,Childress81,lp09}. In this regime, there is a linear relation between the net forces and torques on an immersed flagellum, $(\b{F,N})$, and its associated translational and rotational velocities, $(\bm{U,\omega})$ (rigid body motion is assumed). Consider a rotating helix driving a cell body, as is the case for the swimming runs of flagellated bacteria. In this case, the net forces and torques on the rotating flagellum (or flagella) must balance those of the fluid on the body. Assuming that the cell is axisymmetric about $\b{\hat{x}}$ and swims directly along this axis, we write the body's translational (swimming) velocity as $\b{U}=U\,\b{\hat{x}}$ and its rotational velocity as $\bm{\Omega}=\Omega\, \b{\hat{x}}$. The corresponding fluid force and torque on the cell body are denoted by $\b{F}=-A_0 U\,\b{\hat{x}}$ and $\b{N}=-D_0\Omega\,\b{\hat{x}}$, respectively. A linear mobility relation for the flagellum may then be written as 
\begin{gather}
\left(\begin{array}{cc} \mathcal{A} & \mathcal{B} \\\mathcal{ C} & \mathcal{D} \end{array} \right)\left(\begin{array}{c} U \\ \omega \end{array} \right)=\left(\begin{array}{c} -A_0 U \\ -D_0\Omega \end{array} \right),\label{Eq:FU}
\end{gather}
where we have written the translational and rotational velocities of each point on the flagellum as $\b{U}=U\,\b{\hat{x}}$ and $\bm{\omega}=\omega\, \b{\hat{x}}$, and neglected hydrodynamic interactions between the flagellum and the body. It can be shown that $\mathcal{C}=\mathcal{B}$ \cite{hb65,Purcell97}.  Torque balance requires that the body rotation rate $\Omega$ and the flagellar rotation rate $\omega$ are oppositely signed, so that the cell body counter-rotates with respect to the motion of the flagellum.  The rotary motor at the base of the flagellum attached to the cell body therefore rotates with angular speed $\Omega_m=\omega-\Omega$. 

In order to compare the performance of various polymorphic forms, a hydrodynamic efficiency $\mathcal{E}^*$ is now defined following the work of Purcell \cite{Purcell97}. The power output of the motor, $N\Omega_m$, is compared to the least power that would be required to move the cell body at speed $U$ by any means of propulsion, namely $A_0\,U^2$, and so $\mathcal{E}^*=(A_0\,U^2)/(N\,\Omega_m)$. Expressions for $U$ and $N$ in terms of the rotation rate $\Omega_m$ may be deduced from Eq.~\eqref{Eq:FU}, and various approximations valid for the relative length and velocity scales observed in swimming bacteria may be made (for example $\mathcal{B}^2\ll \mathcal{A} \mathcal{D}$, and $\omega \gg \Omega$) \cite{Purcell97}. Assuming the ability to rescale the propeller dimensions, for a given cell body the maximum value of the swimming efficiency can then be found to be given by $\mathcal{E}=\mathcal{B}^2/(4\mathcal{A}\mathcal{ D})$; $\mathcal{E}$ is the intrinsic propeller efficiency, and is a function of its shape alone \cite{Purcell97,cmyw06}. Note that the  expression for $\mathcal{E}$ could also be reached using a dimensional approach, as $\mathcal{B}$ indicates the correlation between motor torque and forward swimming, while $\mathcal{A}$ and $\mathcal{D}$ are indicative of the fluid friction (via Eq.~\ref{Eq:FU}); the ratio above (or factors thereof) are the only such dimensionally proper arrangements.

To determine the intrinsic efficiency $\mathcal{E}$ for a given experimentally measured or theoretical waveform, we need only compute the three coefficients $\mathcal{A},\mathcal{B}$ and $\mathcal{D}$. To do so accurately, we perform computations using a non-local slender body theory for viscous flows  \cite{kr76,Johnson80}. We consider a single rigid flagellar filament of length $L$ and circular cross-section of radius $\e\, L\, r(s)$, where  $r(s)$ is dimensionless, $\e \ll 1$ is the aspect ratio of the flagellum ($\e \lesssim 10^{-2}$ for bacteria), and $s \in[0,L]$ is the arc-length parameter. For a given translational velocity $U\,\b{\hat{x}}$ and rotational velocity $\omega\,\b{\hat{x}}$ about a point $\b{x_0}$, the fluid force $\b{f}(s)$ on the filament is given implicitly via 
\begin{equation}
8\pi \mu [U\,\b{\hat{x}}+\omega\,\b{\hat{x}}\times(\b{x}(s)-\b{x_0})]=-\b{\Lambda}[\b{f}(s)]-\b{K}[\b{f}(s')](s),\label{sbt}
\end{equation} 
where $\mu$ is the shear viscosity of the fluid, $\b{x}(s)$ denotes the centerline position at a station $s$, and
\begin{gather}
\b{\Lambda}[\b{f}](s)=\left[c(\b{I+\hat{s}\hat{s}})+2(\b{I-\hat{s}\hat{s}})\right]\b{f}(s),\\
\b{K}[\b{f}(s')](s)=\left(\b{I+\hat{s}\hat{s}}\right)\int_{0}^L \frac{\b{f}(s')-\b{f}(s)}{|s'-s|}\,{\rm d}s'+\int_{0}^L \left(\frac{\b{I+\hat{R}\hat{R}}}{|\b{R}(s',s)|}-\frac{\b{I+\hat{s}\hat{s}}}{|s'-s|}\right)\b{f}(s')\,{\rm d}s',
\end{gather}
where $c=-\ln(\e^2 e)$, $\b{R}(s',s)=\b{x}(s')-\b{x}(s)$, $\b{\hat{R}}=\b{R}/|\b{R}|$, $\b{\hat{s}}$ is the local unit tangent vector at the point $s$, and $\b{\hat{s}}\b{\hat{s}}$ is a dyadic product \cite{Johnson80,ts04}.  Henceforth $\b{x_0}$ is set at the origin. In order to obtain numerically the distribution of forces, $\b{f}(s)$, accurately to order $\e^2$, it is required that $r(s)$ decays no slower than $O(\sqrt{s})$ near the filament endpoints, and we have chosen for simplicity $r(s)=\sqrt{4s(L-s)}/L$ as in Ref.~\cite{ts04}. The flagellum diameter $d$ at the midpoint $s=L/2$ is $2\,\e\, L$. The waveforms considered are modeled as perfect helices with centerlines $\b{x}(s)= P\,K\,s \,\b{\hat{x}}+(D/2)\left[\sin(2\pi K s)\,\b{\hat{y}}+\cos(2\pi K s)\,\b{\hat{z}}\right]$, with $K=1/\sqrt{(\pi D)^2+P^2}$, $P$ the pitch, and $D$ the helical diameter.

We solve Eq.~\eqref{sbt}  numerically for $\b{f}(s)$ using a Galerkin method \cite{Atkinson97}, in which $\b{f}(s)$ is written as a finite sum of Legendre polynomials, and Eq.~\eqref{sbt} is required to hold under inner products against the same basis functions. The first integral in the operator $\b{K}[\b{f}]$ is diagonalized in this space \cite{Gotzthesis,ts04}. With $\b{f}(s)$ in hand, we define $F'=\b{\hat{x}}\cdot\int_0^L \b{f}(s)\,{\rm d}s$ and $N'=\b{\hat{x}}\cdot\int_0^L (\b{x}(s)-\b{x_0})\times\b{f}(s)\,{\rm d}s$. Then, setting $(U,\omega)=(1,0)$ we recover $\mathcal{A}=F'$; setting $(U,\omega)=(0,1)$ we recover $\mathcal{B}=F'$ and $\mathcal{D}=N'$. Based on the mathematical accuracy of the method, we estimate that the numerical errors in computing the fluid flow and efficiency calculations for a specified geometry are below $0.1\%$ of the reported values.

\begin{figure}[t]
\begin{center}
\includegraphics[width=4.5in]{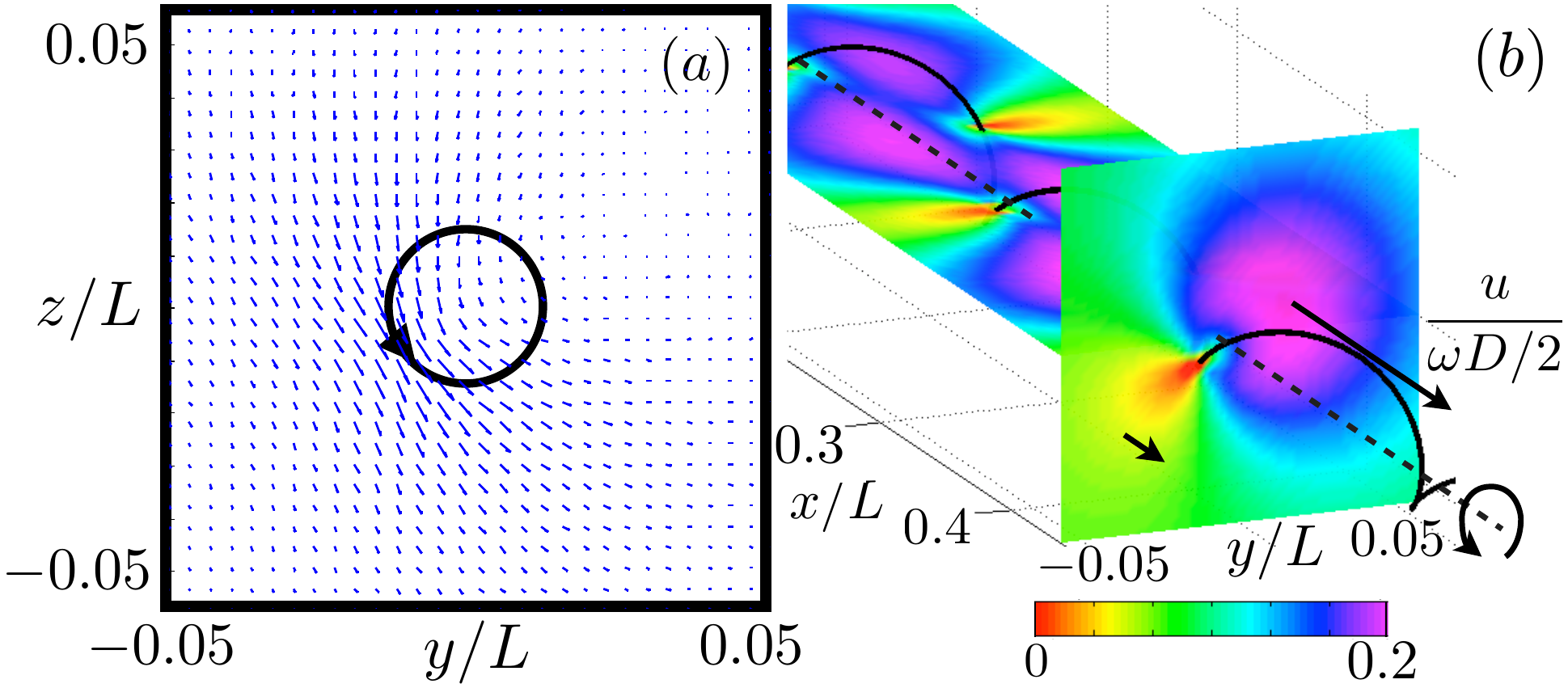}
\caption{FIG. 2 A normal waveform  undergoes pure CCW rotation about the major helical axis $\b{\hat{x}}$, with $(P,D)=(2.3 \,\mu{\rm m},0.4 \,\mu{\rm m})$, $L=10$~$\mu$m, and $d=20$~nm. (a) Velocity vectors through a cross section at $s=L/2$. A dark arrow indicates both the direction of rotation of the flagellar filament, as well as the location on the filament which intersects the cross-sectional plane. (b) The lengthwise velocity $u$ (the fluid velocity through the cross-sectional plane), normalized by the velocity of the flagellum in the cross-sectional plane, $\omega\,D/2$.}
\label{Figure2}
\end{center}
\vspace{-.25in}
\end{figure}

The velocity field, $\b{u}({\bf x})$, at a point $\b{x}$ in the fluid can be recovered using the representation 
\begin{gather}
8\pi \mu\, \b{u(x)}=-\int_{0}^L \left(\frac{\b{I+\hat{R}\hat{R}}}{|\b{R}(s')|}+\frac{\e^2}{2}\frac{\b{I-3\hat{R}\hat{R}}}{|\b{R}(s')|^3}\right)\b{f}(s')\,{\rm d}s',
\end{gather}
where now $\b{R}(s')=\b{x}-\b{x}(s')$ \cite{Gotzthesis,ts04}. We show in Fig.~\ref{Figure2}a  the velocity field so computed through a cross section of a normal flagellar waveform which is undergoing pure CCW  rotation at  rate $\omega$. The flow is primarily restricted to the plane, rotating along with the flagellum (due to the no-slip condition there), and decaying in magnitude away from the intersection point. There is a small lengthwise fluid motion through this plane, so that fluid is slowly shuttled backward along the axis of rotation. This lengthwise velocity $u$ is displayed in Fig.~\ref{Figure2}b, normalized by $\omega D/2$; it is zero at the flagellum boundary (due to the no-slip condition), and increases to a maximum of $u\approx 0.2\, \omega D/2$ on the circular helical perimeter approximately opposite the point where the flagellum intersects the vertical plane.

 \begin{figure*}[t]
\begin{center}
\includegraphics[width=6.3in]{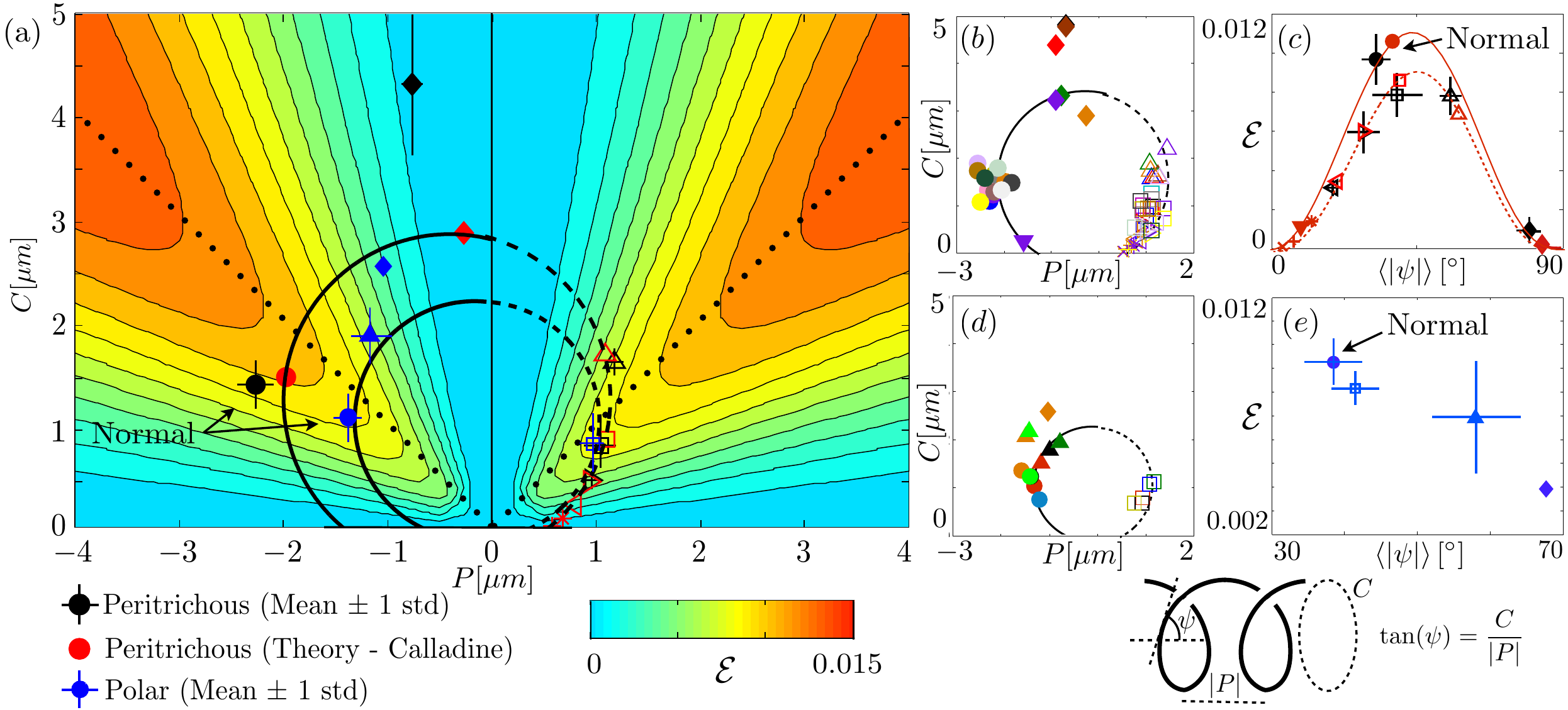}
\caption{FIG 3. (a) Efficiency contours in the circumference-pitch (C-P) plane ($P<0$ for left-handed helices, as in diagram), assuming a flagellum diameter (resp.~length) of $20$~nm (resp.~$10$~$\mu$m), combining the information from Fig.~\ref{Figure3}b-e. The two large circles distinguish the peritrichous and monotrichous (polar) flagellar families, dashed for $P>0$. Data points and bars indicate the mean computed efficiency $\pm$ one standard deviation for the peritrichous (black), polar (blue), and theoretical (red, from Ref.~\cite{Calladine78}) waveforms. Dotted lines indicate the curves $C=\pm P$. (b) Waveform geometries from experimental data for the peritrichous flagella (see Tables S1-S4 in the supplementary material and symbols from Fig.~\ref{Figure1}). Each color represents a different data set. (c) Hydrodynamic efficiencies for each of the peritrichous waveforms as a function of the mean pitch angle, as in (a). Two curves indicate the efficiencies measured continuously along the large circle in the C-P plane in (a); the dashed curve again corresponds to $P>0$ (right-handed helices), and the solid curve to $P<0$ (left-handed helices). (d,e) Same as in (b,c), but for the polar flagellar family \cite{fsa08}. The normal form in each family is the most hydrodynamically efficient of the twelve polymorphic forms by a significant margin.}
\label{Figure3}
\end{center}
\end{figure*}

For each of the experimentally measured waveforms reported in the literature \cite{SuppMat} and the theoretically predicted waveforms \cite{Calladine78}, we compute the intrinsic efficiency $\mathcal{E}$ using the method described above. In each case we assume a flagellum length $L=10$~$\mu$m and diameter $d=20$~nm. Figure~\ref{Figure3}a compiles the efficiency results, further detailed in Fig.~\ref{Figure3}b-e, overlaid upon efficiency contours in the circumference-pitch (C-P) plane ($C=\pi D$). Different symbols represent the various polymorphic forms (see Fig.~\ref{Figure1}), and experimental data show averages  $\pm$ one standard deviation, with peritrichous   (resp. polar) data in  black (resp. blue). As the helix becomes infinitely large (or as the filament becomes infinitesimally slender), Eq.~\eqref{sbt} returns $\mathcal{E}=y^2/(8 y^4+20y^2+8)$, with $y=P/C$. In this limiting case, the efficiency-maximizing geometry has $C=|P|$ (pitch angle $\psi=45^\circ$), indicated in Fig.~\ref{Figure3}a by dotted lines, and $\mathcal{E}=2.8\%$. However, at the biologically relevant length scales and aspect ratio as studied here, for a given pitch $P$ the optimal geometry has $C\approx (7/8)\,|P|$ (pitch angle $\psi\approx 40^\circ$).

We plot in Fig.~\ref{Figure3}b the geometrical data in the (C-P) plane which  allows the different members of the peritrichous flagellar family to be distinguished \cite{fsa08}. Each color represents a different data set \cite{SuppMat}. The mean hydrodynamic efficiencies ($\pm$ one standard deviation) of flagellar polymorphs in the peritrichous family are shown in Fig.~\ref{Figure3}c as a function of the average helical pitch angle, $\langle \psi \rangle$, for measured (black) and theoretically predicted (red) waveforms; the numerical values of the efficiencies for each waveform are noted in the supplementary material. The normal waveform is found to be the most hydrodynamically efficient of the twelve helical forms by a significant margin (with $\langle \mathcal{E} \rangle=0.96\%$) over 23\% more efficient than the next most efficient forms, the curly and semi-coiled waveforms (which are both used by bacteria during change-of-orientation events \cite{Berg2003,db07,dtrb07}). Two curves indicate the efficiencies measured along the large circle in the C-P plane in (a); the larger efficiencies are achieved along this circle when $P<0$. The leftward skew of the theoretical C-P relationship is thus seen to play an important role in the left-handed normal form being more efficient than its right-handed counterparts.

A different flagellar family can be distinguished by examining the circumference-pitch curve for different measurements, and is shown in Fig.~\ref{Figure3}d. These are monotrichous (polar) flagella, assembled from a different flagellin protein than peritrichous flagella, but which follow a similar polymorphic sequence of twelve forms \cite{fsa08}. Similarly to the peritrichous family, the normal form is the most hydrodynamically efficient one (with $\langle\mathcal{E}\rangle=1.03\%$; Fig.~\ref{Figure3}e), a 25\% increase over the next most efficient shape, a right-handed curly waveform.

\begin{figure}[t]
\begin{center}
\includegraphics[width=4in]{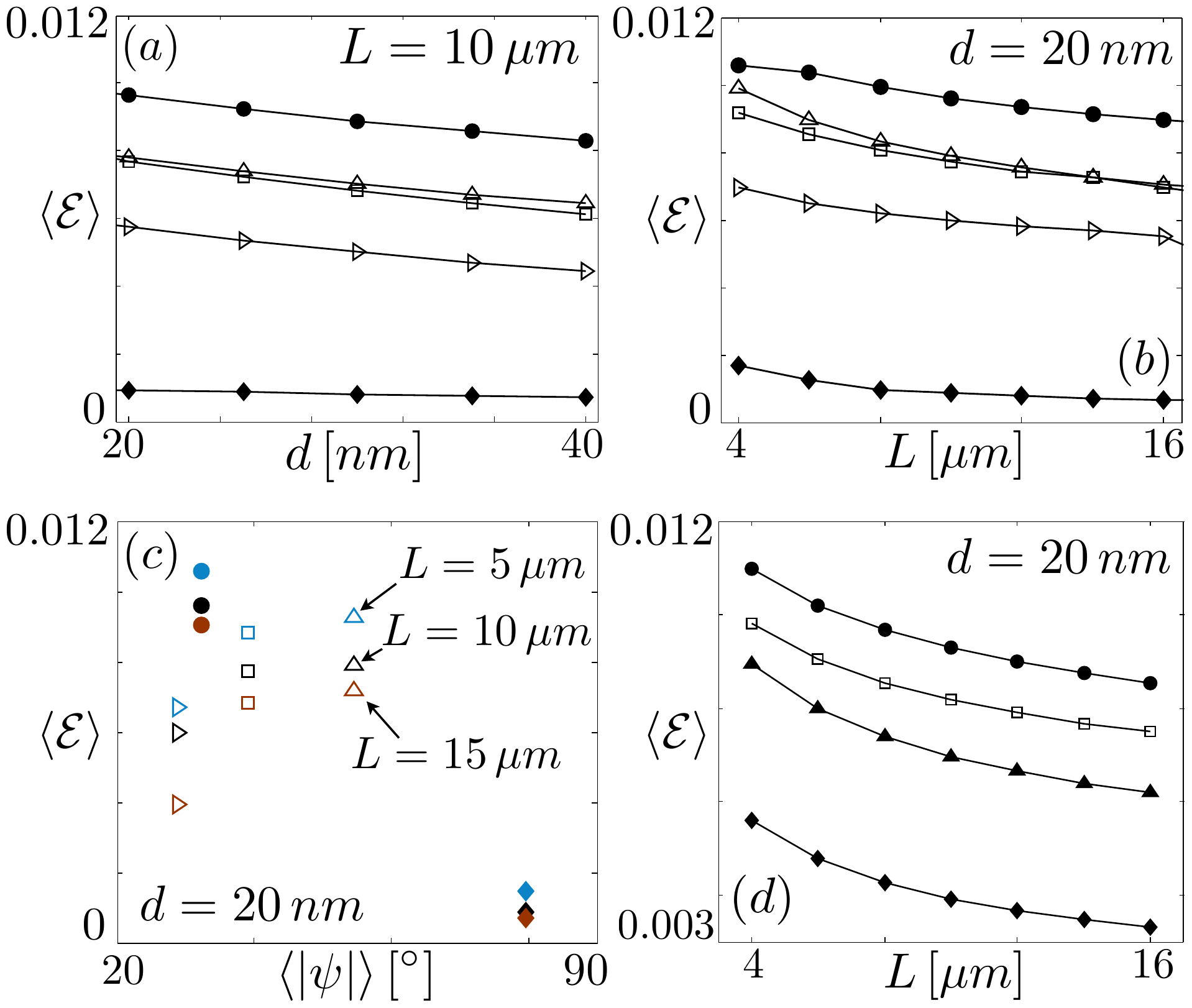}
\caption{FIG. 4 (a) Change in efficiency for the peritrichous family by varying the flagellar diameter from $d= 20$~nm to $40$~nm. 
%could remove sentence
Greater percentage-wise gains in efficiency is obtained for thicker propellers. 
(b) Varying the flagellar length $L$. The efficiency ordering remains nearly the same through the biologically relevant length scales. (c) The mean efficiencies of the normal, coiled, semi-coiled, curly, and curly II waveforms as functions of the mean pitch angle for three different lengths. (d) Same as (b), but for the polar flagellar family.}
\label{Figure4}
\end{center}
\vspace{-.25in}
\end{figure}

To address the robustness of our results against geometrical variations, we changed both the flagellum diameter and length in our computations. We show in  Fig.~\ref{Figure4}a  the mean efficiency computed for the peritrichous waveforms as a function of the flagellum diameter, as a model for the increased effective filament size of flagellar bundles.  The efficiency decreases steadily as the filament size increases, but the efficiency of each polymorph decays at a similar rate, and thus the efficiency ordering from Fig.~\ref{Figure3}c is unchanged. The greatest percentage benefit in efficiency when using the normal form is found when the flagellar diameter is large, e.g. for bundles of many flagella. Varying the flagellar length also shows that the efficiency ordering is not modified, as shown in Figs.~\ref{Figure4}b-c, and the greatest percentage increase in the efficiency of the normal form is achieved for longer filaments. We also changed the lengths used in the computations for the polar flagellar family, with the results shown in Fig.~\ref{Figure4}d, again showing no change in order. For both peritrichous and polar flagellar families, the efficiency orderings shown in Figs.~\ref{Figure3}c-e are therefore robust throughout the biologically relevant parameter space.

Finally, we find that less accurate resistive force theories, which linearly relate body velocities to fluid forces, and are the most widely used approaches for modeling slender bodies in fluids \cite{gh55,Lighthill76}, do not predict the efficiency ordering found using the full non-local hydrodynamics (see also Refs.~\cite{Higdon79,cw09b}) \cite{SuppMat}. Hydrodynamic interactions between different parts of the helical propeller are thus essential in order to conclude on the relative efficiencies of flagellar polymorphs.

In conclusion, by examining all available experimental data on the geometry of bacterial flagella, we found that both peritrichous and monotrichous bacteria employ, among the discrete number of available flagellar shapes, the hydrodynamically optimal polymorph in order to swim in viscous fluids. In contrast to simple estimates showing that locomotion accounts for a negligible portion of a bacterium's metabolic costs \cite{Purcell77}, our results  suggest that fluid mechanical forces may have played a significant role in the evolution of the flagellum

We thank H. C. Berg for discussions, and permission to reproduce the figure from Ref.~\cite{twb00}. We acknowledge the support of the NSF through grant CBET-0746285.
\appendix

\maketitle

\section{Supplementary material}
\subsection{I. Experimental data and computed efficiencies}
Table S1 shows a compilation of measurements from studies on various strains of {\it Salmonella typhimurium}, along with the data sources and the colors used to create Fig.~3b in the main text. Here we have reproduced the measured helical pitch $P$ [$\mu$m] and the helical diameter $D$ [$\mu$m] from the cited sources in the form $(P,D)$. Table S2 contains similar measurements obtained for the organism {\it Escherichia coli}, also included in Fig.~3b. Fujii et al. \cite{fsa08} have considered measurements of a large number of organisms along with their different polymorphic measurements, which we report below as Table S3. These authors have detected different flagellar families corresponding to peritrichous (Family I), monotrichous (or polar) (Family II), lateral (Family III), and some exceptional flagellar filaments; these families are distinguished in the table, and the color schemes match those used to create Figs.~3(b,d). Family I flagellin, Family II flagellin, and Family III flagellin each lead to different circles in the circumference-pitch (C-P) plane, the first two of which are shown in Figs.~3(a,b,d). Table S4 shows the computed data for all the possible theorized waveforms from Calladine's model \cite{Calladine78}, and from a theoretical calculation performed by Hasegawa et al. \cite{hyn98}, which we have included in red in Fig.~3c to show the negligible efficiencies of the more uncommon theoretical polymorphs; note that the unnamed polymorphic forms in Fig.~1a have been obtained in a laboratory setting \cite{kay80}, but are not generally observed in nature. Finally, Table S5 indicates the numerical values of the efficiencies plotted in Figs.~3(c,e).

\renewcommand{\thefigure}{TABLE S1}

\begin{figure}[th]
\begin{center}
\includegraphics[width=5in]{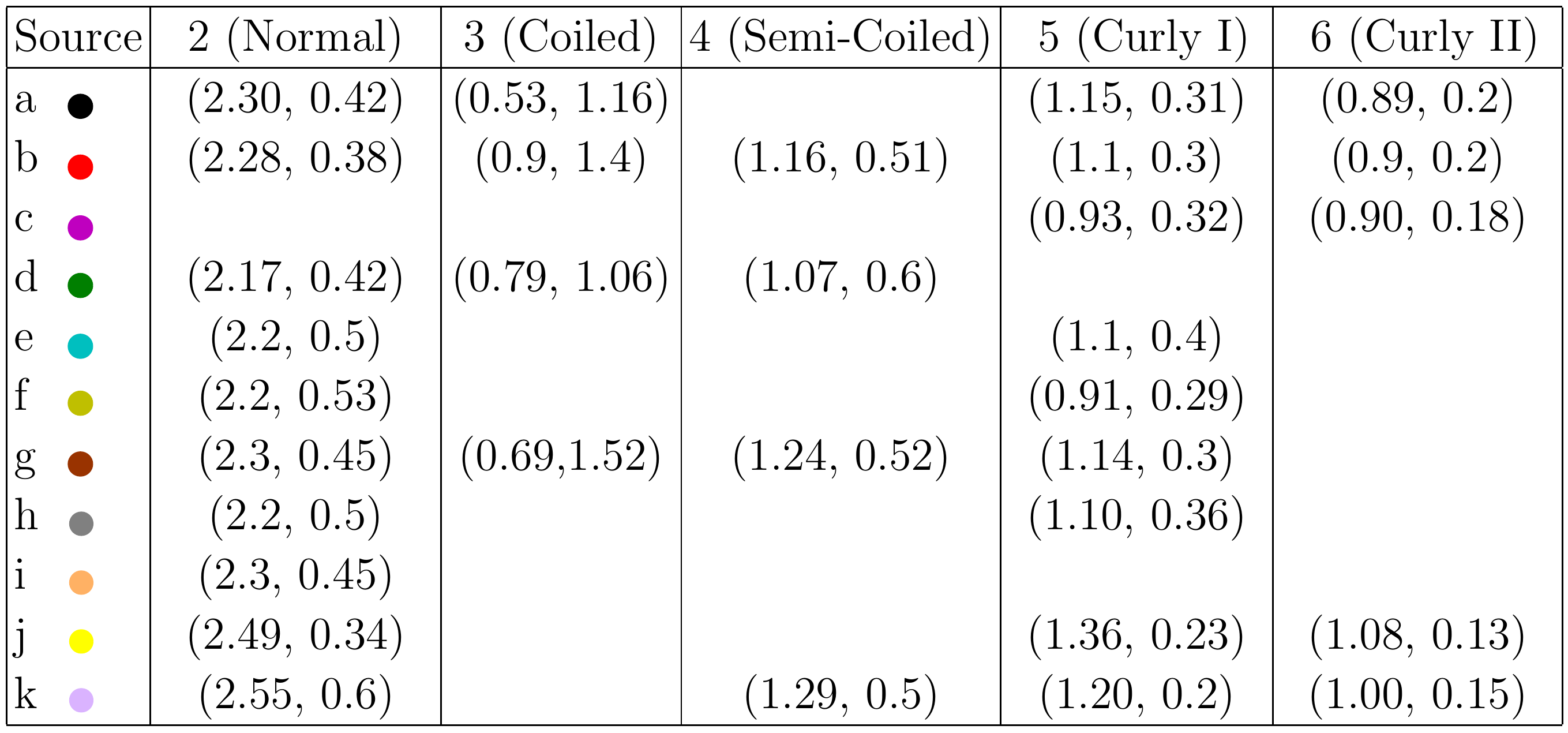}
\caption{TABLE S1. Waveform measurements of the form (helical pitch $P$ [$\mu$m], helical diameter $D$ [$\mu$m]) for {\it Salmonella} from the following sources (organism strain noted in parentheses if reported): a - Kamiya \& Asakura (strain SJ670) \cite{ka76}, b -  Kamiya \& Asakura (SJ25) \cite{ka76}, c - Kamiya \& Asakura (SJ30) \cite{ka76}, d -  Darnton \& Berg (SJW1103) \cite{db07}, e - Iino (SW577) \cite{Iino62}, f - Iino \& Mitani (SJ30) \cite{im66}, g - Hotani \cite{Hotani82}, h - Iino, Oguchi \& Kuroiwa \cite{iok74}, i - Macnab \& Ornston \cite{mo77}, j - Asakura \cite{Asakura70}, k - Fujii, Shabata \& Aizawa \cite{fsa08}. Colors correspond to those in Fig.~3b.}\end{center}
%\label{TableSalmonella}
\end{figure}

\begin{figure}[th]
\begin{center}
\includegraphics[width=4in]{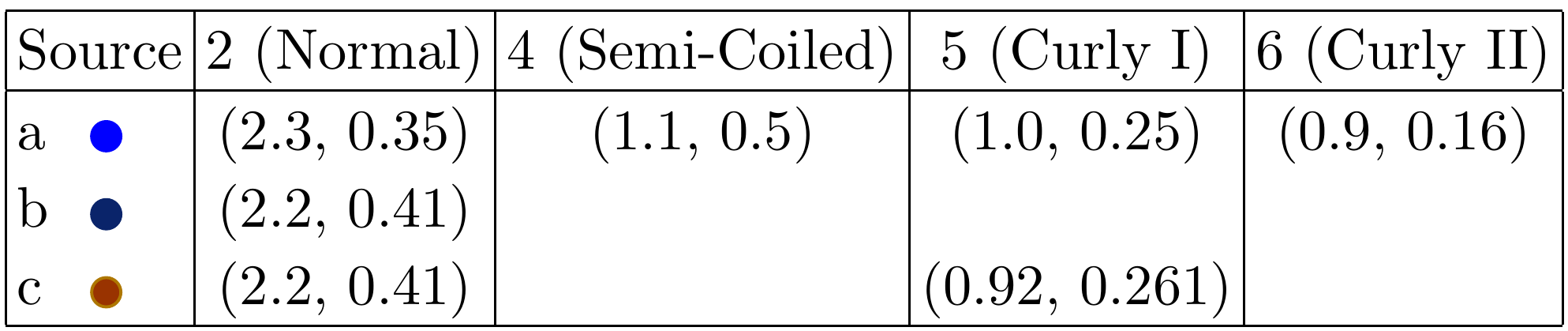}
\caption{TABLE S2. Waveform measurements ($P$ [$\mu$m], $D$ [$\mu$m]) for {\it E. coli.}: a - Turner, Ryu \& Berg \cite{trb00}, b - Matsuura, Kamiya \& Asakura \cite{mka78}, c - Fujii, Shabata \& Aizawa \cite{fsa08}. Colors correspond to those in Fig.~3b.}
%\label{TableEcoii}
\end{center}
\end{figure}

\begin{figure}[th]
\begin{center}
\includegraphics[width=6.1in]{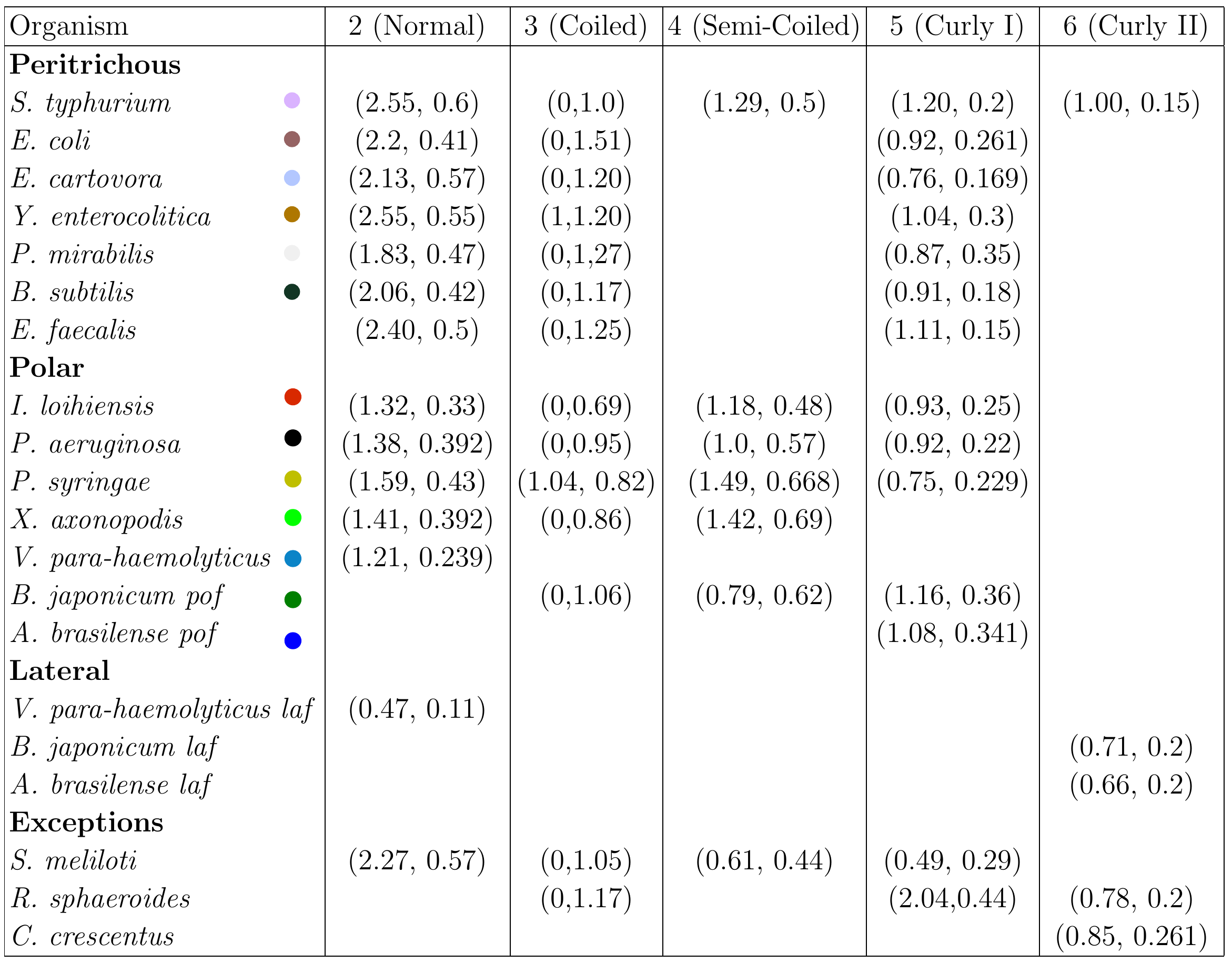}
\caption{TABLE S3. Waveform measurements ($P$ [$\mu$m], $D$ [$\mu$m]) from Fujii, Shibata \& Aizawa \cite{fsa08}, for organisms in the peritrichous, polar, and lateral flagellar families, along with a few exceptions. For polar, lateral, and exceptional flagellar families, the ``Normal'' form refers to small-Normal and very-small-Normal forms (see Ref.~\cite{fsa08}). Colors correspond to those in Fig.~3b (peritrichous) and Fig.~3d (monotrichous, or polar).}
%\label{TableFujii}
\end{center}
\end{figure}

\begin{figure}[th]
\begin{center}
\includegraphics[width=6in]{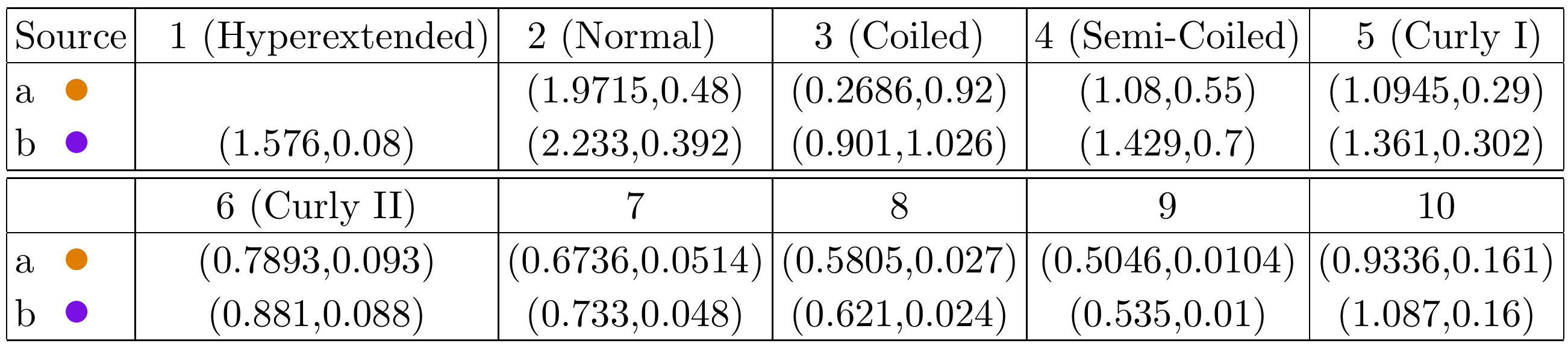}
\caption{TABLE S4. Theoretical waveform data ($P$ [$\mu$m], $D$ [$\mu$m]) from: a - Calladine \cite{Calladine78}, and b - Hasegawa et al. \cite{hyn98}. Colors correspond to those in Fig.~3b.}
%\label{TableModels}
\end{center}
\end{figure}

\begin{table}[th]
\begin{center}
\begin{tabular}{|c|c|c|c|}
\hline
\textbf{Peritrichous} & $\langle \psi \rangle \pm \sigma_{\psi}$ [degrees]&$\langle \mathcal{E} \rangle \pm \sigma_{\mathcal{E}}$ (measured) &\, $ \mathcal{E} $ (theoretical) \cite{Calladine78}\,\\
\hline
1 (Hyper-extended) & $9.1$ &  & $1.1\cdot 10^{-3}$ \\
2 (Normal) & $32.4\pm 4.5$ & $9.6 \cdot 10^{-3}\pm 1.3 \cdot 10^{-3}$ & $1.06\cdot 10^{-2}$ \\
3 (Coiled) & $79.6\pm 3.9$ & $9.3\cdot 10^{-4}\pm 6.9\cdot 10^{-4}$& $7.5\cdot 10^{-4}$\\
4 (Semi-Coiled) & $55.4\pm 3.3$& $7.8\cdot 10^{-3}\pm 1.0 \cdot 10^{-3}$&  $6.9\cdot 10^{-3} $\\
5 (Curly) & $39.0\pm 7.7$& $7.8\cdot 10^{-3}\pm 1.1 \cdot 10^{-3}$& $8.6\cdot 10^{-3} $\\
6 (Curly II) &$ 28.4\pm 5.0$& $5.9\cdot 10^{-3}\pm 1.0\cdot 10^{-3}$& $6.0\cdot 10^{-3}$\\
7 & $20.3$ & &  $3.4\cdot 10^{-3}$ \\
8&  $13.5$  & &  $1.4\cdot 10^{-3}$  \\
9& $8.3$ & & $ 3.4\cdot 10^{-4} $ \\
10 & $ 3.7$ & & $5.1\cdot 10^{-5}$ \\
\hline
\textbf{Polar} & &  & \\
\hline
(Normal) & $38.6\pm 4.0$ & $1.03 \cdot 10^{-2}\pm 1.0 \cdot 10^{-3}$& \\
(Coiled) & $68.0$ & $3.9\cdot 10^{-3}$& \\
(Semi-Coiled) & $58.4\pm 6.3$& $7.0\cdot 10^{-3}\pm 2.3 \cdot 10^{-3}$& \\
(Curly) & $41.7\pm 3.2$& $8.2\cdot 10^{-3}\pm 0.7 \cdot 10^{-3}$& \\
\hline
\end{tabular}
\end{center}
\caption{TABLE S5. Mean pitch angles and efficiencies $\pm$ one standard deviation (when available) for the peritrichous flagellar family (as in Fig.~3c) for measured and theoretical waveforms \cite{Calladine78}, and for measured waveforms from the polar (or monotrichous) family (as in Fig.~3e).}
%\label{ETable}
\end{table}
\clearpage
\subsection{II. Waveform geometries and resistive force theory predictions}

The primary results found from comparing the hydrodynamic efficiencies of the polymorphic forms were reported in the main text. Most notably, the normal polymorphic form was found to be the most efficient waveform by a significant margin for both peritrichous and monotrichous (polar) flagellar families. Figures~3(b,d) showed the geometries of the waveforms considered in the helical circumference-pitch (C-P) plane (with $C=\pi D)$. Here we provide a different standpoint from which to visualize the geometries; Fig.~S1 shows the geometrical relations in pitch angle $\psi$ vs. circumference $C$ for the peritrichous and monotrichous (polar) flagellar families. The normal, semi-coiled, and curly forms all occupy nearby regions of parameter space in pitch angle $\psi$. However, compared to the other polymorphs, the normal waveforms have a significantly larger helical circumference for a given pitch angle.

\begin{figure}[htbp]
\begin{center}
\includegraphics[width=5in]{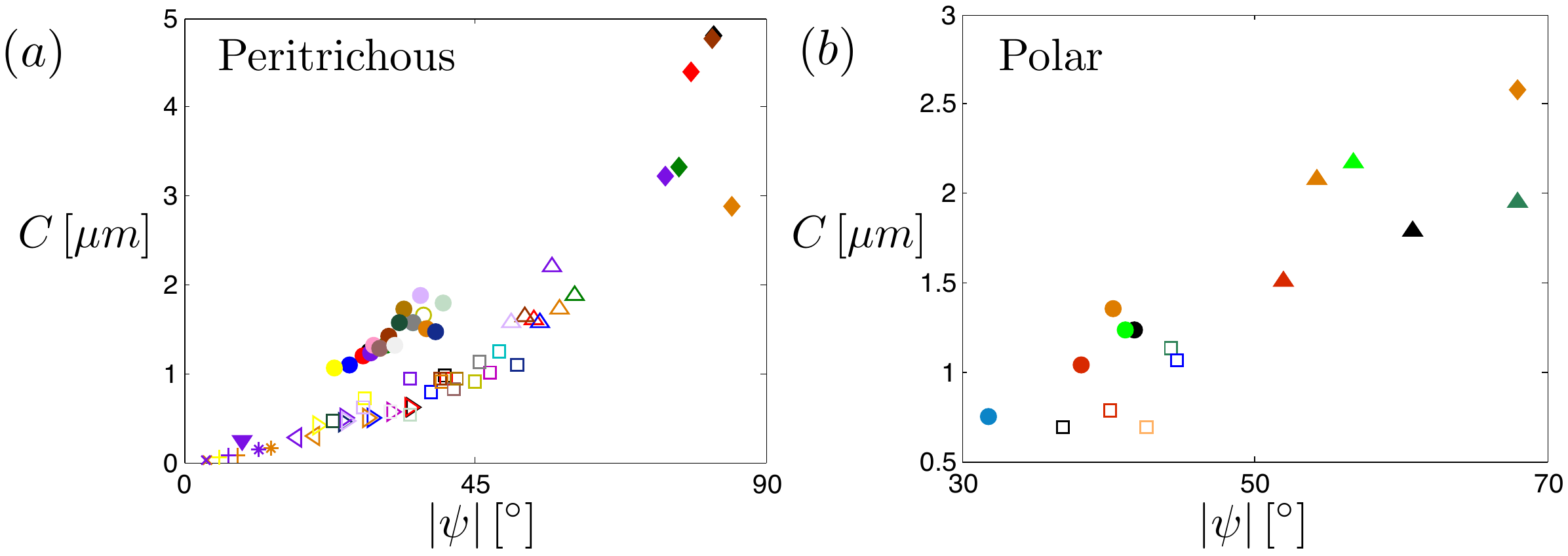}
\caption{FIG. S1. Geometrical data for the (a) peritrichous and (b) monotrichous polar flagellar families (see Fig.~1a for symbol legend).  The normal, semi-coiled, and curly forms all occupy nearby regions of parameter space in pitch angle $\psi$. However, compared to the other polymorphs, the normal waveforms have a significantly larger helical circumference for a given pitch angle.}
%\label{PeritrichousPsiC}
\end{center}
\end{figure}
\begin{figure}[t]
\begin{center}
\includegraphics[width=5in]{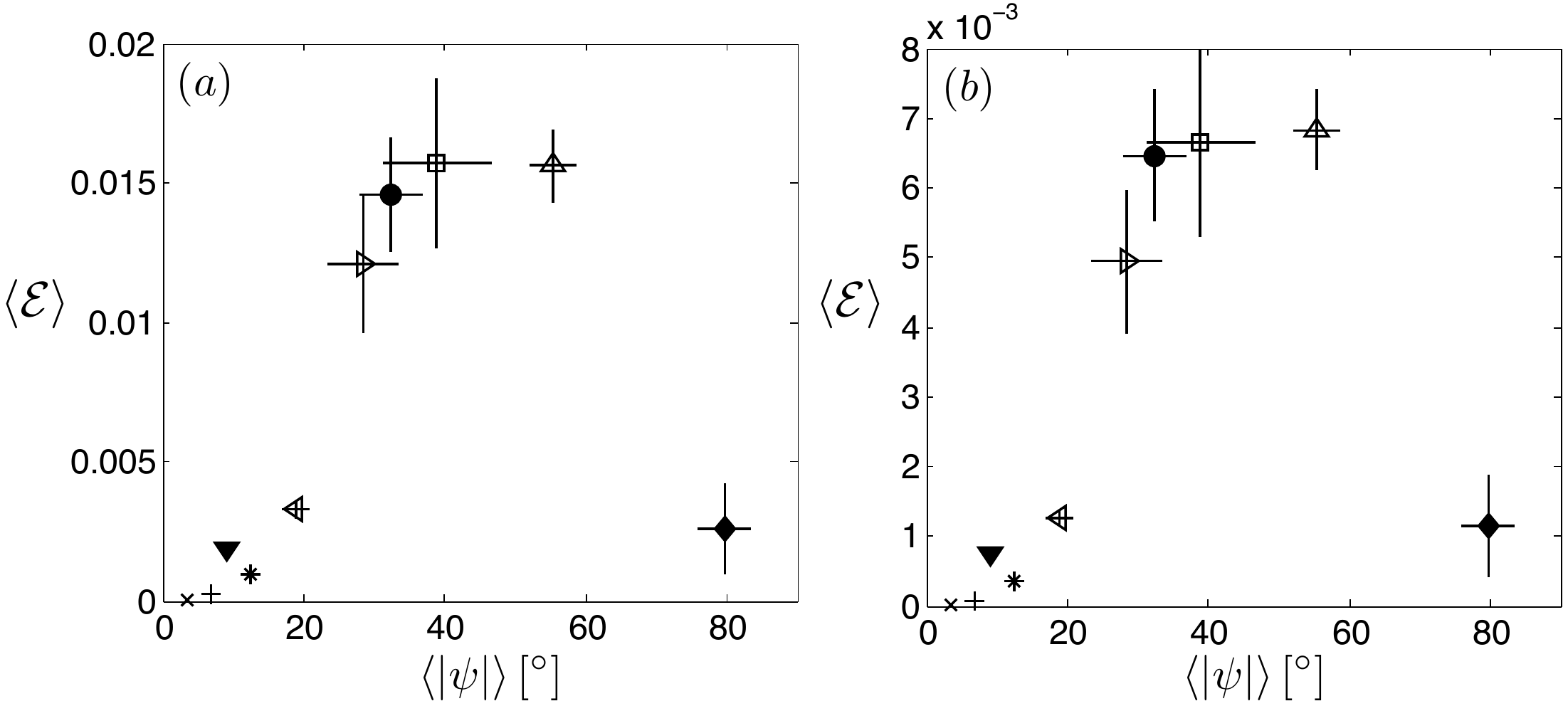}
\caption{FIG. S2. Mean efficiencies computed using two local resistive force models; namely, (a) the local theory achieved by simply neglecting the non-local term $\b{K}[\b{f}(s')](s)$ in Eq.~(2) in the main text, and (b) the local theory of Lighthill for helical waveforms \cite{Lighthill75}. Neither local theory predicts the correct efficiency ordering of the polymorphic forms for biologically relevant parameters.}
%\label{RFTExamples}
\end{center}
\end{figure}

Finally, resistive force theories,  which are valid only at $O(\log 1/\e )^{-1}$ and relate local body velocities to local fluid forces, are the most widely used approaches for modeling slender bodies in fluids, dating in the case of highly viscous flow back to the seminal work of Gray \& Hancock \cite{gh55}. However, we note that the local resistive force theory achieved by ignoring the non-local integral operator $\b{K}[\b{f}(s')](s)$ in Eq.~(2) in the main text (see also Ref.~\cite{gh55}), and even the more appropriate resistive force theory for helical geometries due to Lighthill \cite{Lighthill76} do not predict the efficiency ordering found using the full non-local slender body theory. A related study by Chattopadhyay and Wu also suggests the importance of solving for the full nonlocal fluid interactions in such systems \cite{cw09b}. Figures~S2(a,b) show the efficiencies computed using these local theories for the peritrichous flagellar family data. The first approximation significantly overestimates the efficiencies, and the curly and semi-coiled forms are the most efficient. The second approximation (using Lighthill's resistive coefficients) significantly underestimate the efficiencies, and again the curly and semi-coiled forms are computed to be the most efficient. 
Hydrodynamic interactions between different parts of the flagella, which are captured by our slender-body approach but not in resistive force theory, are thus essential in order to conclude on the relative efficiencies of flagellar polymorphs.

\bibliographystyle{apsrev}
\bibliography{BigBib}

\end{document}